\documentclass[twocolumn,showpacs,preprintnumbers,amsmath,amssymb]{revtex4}

\usepackage{graphicx}
\usepackage{dcolumn}
\usepackage{bm}

\begin{document}


\title{Observation of the Isotopic Evolution of Pressurized Water Reactor Fuel Using an Antineutrino Detector}

\author{N.~S.~Bowden}
\affiliation{Sandia National Laboratories, 7011 East Ave, Livermore,
CA 94550 }
\altaffiliation{Current Address: Lawrence Livermore National Laboratory, 7000 East Ave., Livermore,
CA 94550}
 \email{nbowden@llnl.gov}

\author{A.~Bernstein}
\author{S.~Dazeley}
\affiliation{Lawrence Livermore National Laboratory, 7000 East Ave., Livermore,
CA 94550}

\author{R.~Svoboda}
\affiliation{Lawrence Livermore National Laboratory, 7000 East Ave., Livermore,
CA 94550}
\affiliation{Department of Physics, University of California, One Shields Ave, Davis, CA 95616}

\author{A.~Misner}
\author{T.~Palmer}
\affiliation{Department of Nuclear Engineering, Oregon State University, Corvallis, OR 97331}

\date{\today}

\begin{abstract}
By operating an antineutrino detector of simple design during several fuel cycles, we have observed long term changes in antineutrino flux that result from the isotopic evolution of a commercial Pressurized Water Reactor (PWR). Measurements made with simple antineutrino detectors of this kind offer an alternative means for verifying fissile inventories at reactors, as part of International Atomic Energy Agency (IAEA) and other reactor safeguards regimes.

\end{abstract}

\maketitle

\section{Introduction}
\label{sec:intro}

In the $50$~years since antineutrinos were first detected using a nuclear reactor as the source~\cite{reines}, these facilities have played host to a large number of neutrino physics experiments. About two decades later it was realized~\cite{nu77} that the techniques and technologies developed for the study of neutrino physics could also be applied to nuclear reactor monitoring in the context of nuclear safeguards. This capability was demonstrated by several later physics experiments~\cite{rovno,bugey}. More recently it has been pointed out that the field of antineutrino detection has matured sufficiently to envisage the deployment of practical devices dedicated to reactor safeguards~\cite{firstpaper}.

Safeguards agencies, such as the IAEA, use an ensemble of procedures and technologies to detect diversion of fissile materials from civil nuclear fuel cycle facilities into weapons programs. Nuclear reactors generate up to several hundreds of kilograms of plutonium per year, and are thus an especially important part of this cycle, as it is here that plutonium is produced. Current safeguards practice at reactors is focused upon tracking fuel assemblies through item accountancy and surveillance, and does not typically include direct measurements of fissile inventory. While containment and surveillance practices are considered effective, they are also costly and time consuming for both the agency and the reactor operator.

In this paper, we describe how we observe the effect of fuel evolution upon the rate at which the reactor emits antineutrinos. This data was acquired using a simple, robust, and nonintrusive detector, ``SONGS1''~\cite{secondpaper,powerpaper}, which we have installed at a commercial nuclear power reactor. Observing this evolution with an antineutrino detector allows limits to be placed upon the fissile inventory of the reactor, in particular the amount of $^{235}$U consumed and $^{239}$Pu produced.

\section{Antineutrino Production in Reactors}

\subsection{Antineutrino Production Rate}
\label{sec:nubar_production}

Antineutrino emission by nuclear reactors arises from the beta decay of neutron-rich fragments produced in heavy element fissions. On average, fission is followed by the production of approximately six antineutrinos, which emerge from the core isotropically and effectively without attenuation. Furthermore, the average number of detectable antineutrinos produced per fission is significantly different for the two major fissile elements, $^{235}$U and $^{239}$Pu. Hence, as the core evolves and the relative mass fractions and fission rates of $^{235}$U and $^{239}$Pu change, the number of detected antineutrinos will also change. This relation between the mass fractions of fissile isotopes and the detectable antineutrino flux is known as the burnup effect, and has been observed consistently in previous neutrino physics experiments \cite{reactor_nu_review}.

\begin{figure}[tb]
\centering
\includegraphics*[width=3in]{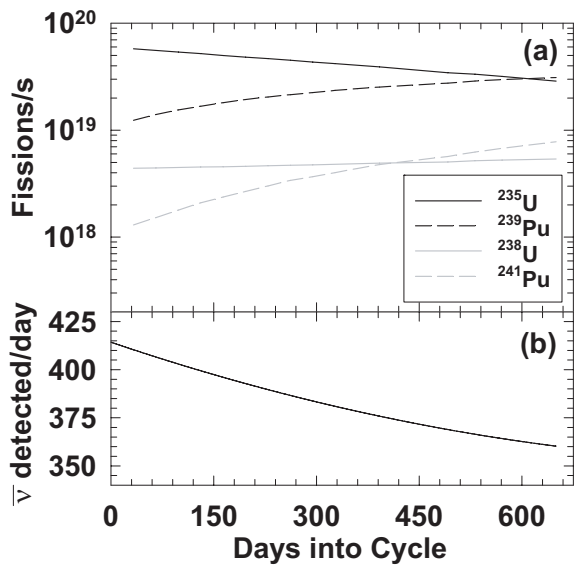}
\caption{The predicted (a) fission rates for the four main fissioning isotopes
and (b) antineutrino detection rate in SONGS1 throughout a $650$~day reactor equilibrium fuel cycle.} \label{fig:fisrates}
\end{figure}

To evaluate the antineutrino production rate we begin by considering the key operational parameter of a reactor, the thermal power $P_{th}$;
\begin{equation}
P_{th} = \sum_{i} F_{i} \cdot E^{f}_{i}, \label{eq:power}
\end{equation}
where $F_{i}$ is the fission rate for isotope $i$ and $E^{f}_{i}$ is the average energy released per fission for this isotope. The sum runs over all fissioning isotopes. However, in practice it is sufficient to consider only the four dominant fissioning isotopes $^{235}$U, $^{238}$U, $^{239}$Pu and $^{241}$Pu, which account for $\approx 99.9\%$ of fissions in a typical reactor.

Following the formulation of \cite{huber}, it is useful to define the power fractions contributed by each isotope as:
\begin{equation}
f_{i} =\frac{F_{i}\cdot E^{f}_{i}}{P_{th}}. \label{eq:powerfracs}
\end{equation}
Even with the reactor operated at constant thermal power these fractions are time-dependent, changing by several tens of percent throughout a typical cycle as, for example, $^{235}$U is consumed and $^{239}$Pu is produced by neutron capture on $^{238}$U.

Using tabulated values of the antineutrino production number densities~\cite{huber} for each fissioning isotope, we can express the total antineutrino emission rate, $n_{\bar{\nu}}$, in terms of the thermal power and the power fractions:
\begin{equation}
n_{\bar{\nu}}(t)  = P_{th}(t) \sum_{i} {\frac{f_{i}(t)}{E^{f}_{i}}} \int
dE_{\bar{\nu}} \phi_{i}(E_{\bar{\nu}}) , \label{eq:nu_p_rate}
\end{equation}
where the explicit time dependence of the power fractions and, possibly, the thermal power are noted. Here, $\phi(E_{\bar{\nu}})$, is the antineutrino energy dependent number density per MeV and fission for the $i$th isotope.

\subsection{Antineutrino Detection Rate}
\label{sec:nubar_production}

Antineutrinos are detected via the inverse beta decay process on quasi-free protons in hydrogenous scintillator: $\bar{\nu}_e + p \rightarrow e^{+} + n$. The number of antineutrinos produced per unit energy varies with isotope above the $1.8$~MeV threshold for this reaction, so that the antineutrino \textit{detection} rate will also vary as the fuel composition of a reactor evolves. This relation between fissile plutonium and uranium content and the antineutrino detection rate provides a means by which to use these particles to provide information relevant to reactor safeguards.

To derive the antineutrino detection rate $N_{\bar{\nu}}$, we must fold the energy dependent inverse beta cross section, $\sigma(E_{\bar{\nu}})$, and detector efficiency, $\epsilon(E_{\bar{\nu}})$, into the integral of Eq.~\ref{eq:nu_p_rate}:
\begin{equation}
N_{\bar{\nu}}(t)  = \left(\frac{N_{p}}{4 \pi D^{2}}\right) P_{th}(t) \sum_{i}
{\frac{f_{i}(t)}{E^{f}_{i}}} \int dE_{\bar{\nu}} \sigma \phi_{i} \epsilon,
\label{eq:nu_d_rate}
\end{equation}
The prefactor includes $N_{p}$, the number of target protons in the active volume of the detector, and $D$, the distance from the detector to the center of the reactor core.

Following the formulation of \cite{rovno}, it is instructive to rewrite
Eq.~\ref{eq:nu_d_rate} as:
\begin{equation}
N_{\bar{\nu}}(t)  = \gamma \left(1+k(t)\right) P_{th}(t), \label{eq:nu_d_rate2}
\end{equation}
where $\gamma$ is a constant encompassing all non varying terms (e.g. detector size, detector/core geometry), and $k(t)$ describes the change in the antineutrino flux due to changes in the reactor fuel composition. $\gamma$ is chosen so that the value of $k$ at the beginning of a reactor fuel cycle is zero.

Typically, commercial reactors are operated at constant thermal power, independent of the ratio of fission rates from each isotope. Operating in this mode, the factor $k$ -- and therefore the antineutrino detection rate $N_{\bar{\nu}}$-- decreases by about $0.1$ over the course of a reactor fuel cycle, depending on the initial fuel loading and operating history. The magnitude of this burnup effect can be predicted at the few percent level in an absolute sense, if the reactor fuel loading and power history are known. Much of the uncertainty arises from systematic shifts in measured antineutrino energy densities $\phi_{i}(E_{\bar{\nu}})$, so that the relative uncertainty in the predicted burnup rate can be considerably smaller.

\subsection{Reactor Simulation and Fission and Antineutrino Detection Rate Prediction}
\label{sec:reactorsim}

In order to predict $k(t)$, and thus the evolution of the detected antineutrino rate, for the SONGS Unit~2 reactor we used the ORIGEN/SCALE simulation package~\cite{origen} to predict the fission rates per isotope throughout a typical reactor fuel cycle. At discrete steps, the simulation returns the mass, fission rate, and number densities of fissile isotopes in each fuel assembly of the core. The per assembly fission rates are summed over the core to produce an estimate of the reactor evolution history (Fig.~\ref{fig:fisrates}a).

The input fuel isotopics for this simulation are taken from the San Onofre Nuclear Generating Station Final Safety Analysis Report (FSAR) \cite{FSAR}. This report describes a nominal fuel loading which differs by a few percent from the actual fuel load in recent cycles. While this difference will lead to a systematic shift in our predicted count rate from the true count rate, the FSAR isotopics are sufficient for comparison with our current data, and have the important advantage of being publicly available and not subject to use restrictions by the utility.

To predict the antineutrino detection rate, the fission rates are folded with measured or estimated antineutrino energy densities, the inverse beta interaction cross-section, and an estimate of the detector response function. The antineutrino energy densities are provided from $1.8~$MeV to $10~$MeV in $0.5~$MeV steps in \cite{vogel1} and \cite{vogel2}. Various parameterizations exist for this data \cite{schreckenbach,huber}. In our parametrization, we used a cubic interpolation between adjacent data points to extract the densities and errors at intervening energies.

We predict a change in the detected antineutrino rate of about $0.5$\%/month (Fig.~\ref{fig:fisrates}b) for SONGS Unit~2, and a thus a decrease in rate of 10\% over the course of the entire fuel cycle. Similarly, we expect a step increase in the detection rate of this amount after reactor refueling, when spent fuel laden with $^{239}$Pu is removed and replaced with fuel containing only $^{235}$U and $^{238}$U.

\section{Antineutrino rate measurements for fissile inventory safeguards}
\label{sec:safeguards}

As shown in Eq.~\ref{eq:nu_d_rate2}, the time dependence of the antineutrino detection rate also depends on the evolution of both the thermal power and the core isotopics. As a consequence, an independent measurement of $P_{th}(t)$ is required if antineutrino detection rate measurements alone are used to constrain reactor fissile inventory. In this case, the inspector would require access to the reactor power history.

Used in concert with operator declarations, antineutrino rate measurements may be of considerable safeguards value. Declarations of fresh and spent fuel and operating history, whose self-consistency is already checked as part of the IAEA regime, must now in addition be consistent with the slow change in antineutrino rate throughout each cycle (``slope'' changes) and the sharp increase in rate observed following refueling  (``step'' changes). Moreover, for reactors which have proceeded through several refueling cycles, a standard operating history emerges, so that a comparison of slope and step \textit{between} cycles offers additional confirmation of standard operation. This latter type of comparison is slightly different than verification within a single cycle, inasmuch as it is less dependent on year-to-year operation declarations.

It is also interesting to note that the detected antineutrino energy spectrum is different for the two main fissioning isotopes. This means that a completely independent estimate of fissile content can be derived from the antineutrino \textit{energy spectrum} without relying on operator declarations~\cite{huber}.

The requirements for a detector capable of making such a spectral measurement would be considerably more stringent than is the case for a rate measurement. For example, excellent counting statistics would be required in the high energy region ($7-9$~MeV), necessitating a more efficient and/or larger detector, as well a good energy calibration across the entire energy range. We expect it would be difficult to design a detector with these additional features, while still preserving the operational simplicity and low cost of our current design.

\section{The SONGS1 Detector}
\label{sec:detector}

\begin{figure}[tb]
\centering
\includegraphics*[width=3in]{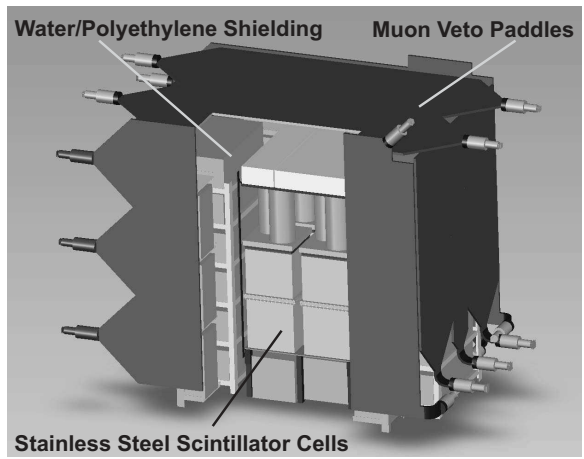}
\caption{A cut away diagram of the detector, showing the major subsystems.}
\label{fig:detector}
\end{figure}

The SONGS1 detector (Fig.~\ref{fig:detector}) has been described in detail elsewhere~\cite{secondpaper}. SONGS1 comprises a central target region, a $\approx 0.5$~m thick layer of passive shielding, and a plastic scintillator muon veto. The target comprises $\approx 0.64$~tons of gadolinium (Gd) loaded liquid scintillator~\cite{paloverde_scint} contained in $4$~stainless steel cells each of which is readout by two Photo Multiplier Tubes (PMTs).

The positron ($e^{+}$) and neutron ($n$) produced by the inverse beta interaction are detected in close time coincidence, allowing strong rejection of the much more frequent single event backgrounds due to natural radioactivity or cosmic ray backgrounds. The incorporation of Gd in the liquid scintillator greatly aids this rejection by decreasing the neutron capture time to $28~\mu$s, compared to $\approx 200~\mu$s for capture on hydrogen.

Unfortunately, this reaction is not the only means by which correlated pairs of interactions can be produced in the scintillator. For example, cosmic ray muons can produce fast neutrons in the material surrounding the detector -- these neutrons can slow due to elastic collisions with protons, whose recoils are detected by the scintillator, and then capture on Gd with the characteristic $28~\mu$s time constant. Such processes are referred to as correlated backgrounds. It is for this reason that SONGS1 incorporates a muon veto system. The random coincidence of two unrelated single background interactions (gamma rays or neutrons) can also occur within the short time characteristic of neutron captures on Gd. The statistical rejection of these \textit{uncorrelated backgrounds} is described in Sec.~\ref{sec:analysis}, while the passive shielding incorporated into SONGS1 reduces the frequency with which these occur.

A Data Acquisition System (DAQ) based upon standard NIM and VME electronics modules records timing and PMT amplitude information. A trigger is generated when the signals from the two PMTs on a cell exceed an $\approx 1.5$~MeV threshold and no hit was recorded in the muon veto in the preceding $20~\mu$s. On a trigger, ADCs record the amplitude of all $8$~PMTs and Time-to-Amplitude Converter (TAC) pulses corresponding to the time to last trigger (interevent time) and the time since last hit in the muon veto (or, because of hardware limitations, the time since the last trigger, whichever is less).

We estimate the detection efficiency of the complete SONGS1 detector system, including the DAQ, to be $10.7 \pm 1.5 \%$. When combined with the expected antineutrino interaction rate ($3800 \pm 440$/day) at the beginning of cycle, this yields an expected antineutrino detection rate of $407\pm75$~/day.

SONGS1 is located in the tendon gallery of the Unit~2 reactor at the San Onofre Nuclear Generating Station (SONGS). This is an ideal location, since it places the detector as close to the reactor core as possible ($24.5\pm1$~m) while remaining outside containment, away from regular plant operations, and about $10$~m below the surface which attenuates correlated background generating cosmic ray muons.


\section{Data Analysis}

\subsection{Antineutrino Event Selection} \label{sec:analysis}

Again, the details of the event selection have been described in detail elsewhere~\cite{secondpaper}. The analysis begins by forming consecutive triggers streamed to disk by the DAQ into events pairs (the two halves of which are denoted ``prompt'' and ``delayed''). The raw data thus formed comprises prompt and delayed PMT amplitudes, the time between the prompt and delayed events, and the time between the prompt event and the last muon to traverse the veto.

An online calibration is performed using a spectral feature due to the $2.6~$MeV $^{208}$Tl gamma ray from the thorium chain, which originates from the material surrounding the detector. In our relatively small detector, this feature will be shifted lower in energy due to escape of some of the gamma energy. We have used an MCNPX \cite{MCNPX} simulation to estimate this shift as $0.21$~MeV, such that the $2.6$~MeV peak would appear at $2.39$~MeV. Setting the analysis threshold at this peak in the energy spectrum reduces the impact of calibration uncertainties, since no extrapolation in the energy scale is needed.

Table~\ref{tab:cuts} lists the selection cuts that are then applied to extract candidate antineutrino events (correlated events). These cuts are applied to each scintillator cell separately so that each is considered an independent detector. While this results in an $\approx 10$\% loss of detector efficiency, it allows for a more simple and flexible analysis. A brief description of each cut is given below:

\begin{itemize}

\begin{item}
\textit{PMT ratio cut}: A cut on the ratio of the light seen by the two PMTs on a cell,
\begin{equation}
r = \left|\frac{PMT_A-PMT_B}{PMT_A+PMT_B}\right| < 0.4,
\end{equation}
is enforced. This is to ensure that energy depositions that occur much closer to one PMT than the other, and that hence have an atypically large light collection efficiency, are excluded.
\end{item}

\begin{item}
\textit{Prompt energy cut}: The prompt energy threshold is placed at the calibration point, $2.39~$MeV, so that no extrapolation in energy is required. No events are admitted beyond $9~$MeV as it is expected, and observed, that there are very few antineutrino interactions above this energy.
\end{item}
\begin{item}
\textit{Delayed energy cut}: The delayed energy threshold is placed at $3.50$~MeV. This threshold is optimized to keep the number of uncorrelated events as low as possible, which requires a high threshold, while still staying as close as possible to the $2.39$~MeV calibration peak to reduce calibration uncertainties. Events with delayed energy of greater than $10~$MeV are also excluded, as the predicted delayed energy spectrum is zero beyond this energy.
\end{item}
\begin{item}
\textit{Time since last muon cut}: Event pairs closer in time to a muon than $100~\mu$s are excluded, since these have a high likelihood of having been produced by the muon.
\end{item}
\end{itemize}

\begin{figure}[tb]
\centering
\includegraphics*[width=3in]{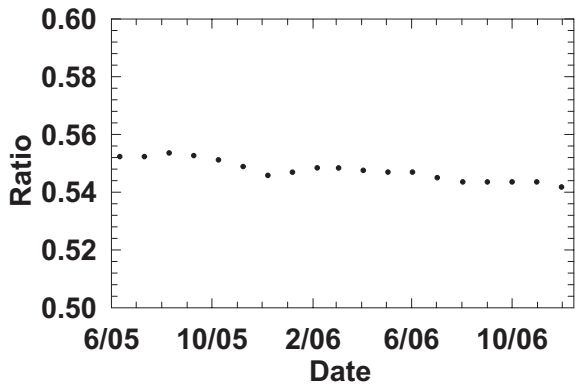}
\caption{The fraction of the prompt events used for calibration that also pass the more strict antineutrino analysis cut (calibration employs a PMT ratio cut of $0.4$ while analysis employs $0.1$).
The variation in this fraction over the $18$~month data taking period is less than $2$\%.} \label{fig:ratio_stab}
\end{figure}

It should be noted that we apply a stricter PMT ratio cut (less than $0.1$) when performing the online calibration. While this samples a smaller volume of the cell than the general event selection, it improves the reliability with which the calibration peak can be found. We are aware of no effect which alters the relative calibrations between these two volumes. For example, examination of the relative number of events passing the prompt and delayed energy cuts with these two ratio values (Fig.~\ref{fig:ratio_stab}) shows little time dependent variation.

\begin{table}
\caption{List of selection cuts applied to extract candidate antineutrino events.} \label{tab:cuts}
\begin{tabular}{l r r} \hline
Cut       & ~~~Lower Limit               & ~~~~Upper Limit \\ \hline
PMT ratio &~-~        &  ~$0.4$      \\
Prompt Energy   &~$2.39$~MeV        &  ~$9$~MeV      \\
Delayed Energy    &~$3.50$~MeV           &  ~$10$~MeV         \\
Time since last muon   &~$100~\mu$s          &           \\
\hline
\end{tabular}
\end{table}

\begin{figure}[tb]
\centering
\includegraphics*[width=3in]{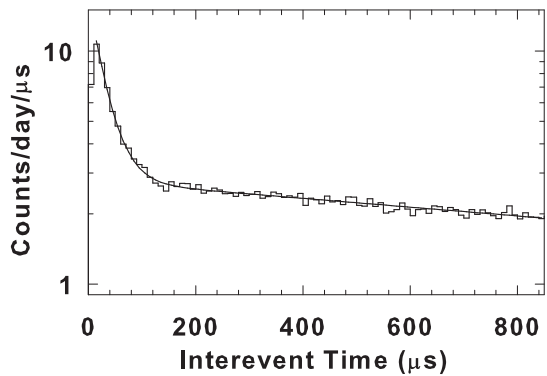}
\caption{A representative interevent time spectrum, displaying the two expected exponential features associated with correlated and uncorrelated events.} \label{fig:interevent}
\end{figure}

Next, an interevent time spectrum is populated using events that pass all selection cuts (Fig~\ref{fig:interevent}). Two clear exponential features are seen, one due to correlated events with a time constant equal to the neutron capture time of the Gd loaded liquid scintillator ($\approx 28~\mu$s) and the other due to accidental coincidences between two successive uncorrelated events, with time constant equal to the inverse of the detector trigger rate. Event by event these two classes cannot be distinguished, so we employ a statistical separation, fitting the interevent time spectrum to the sum of two exponentials. Simple exponential integrals then yield the number of events in each class.

For this analysis we consider data acquired during an $18$~month period beginning in June of $2005$ and ending in November of $2006$. This period encompasses the last $7$~months of the SONGS Unit~$2$ Fuel Cycle~$13$, a $3.5$~month refueling outage, and the first $6$~months of the SONGS Unit~$2$ Fuel Cycle~$14$. We choose an integration time of $30$ days per data point, since we expect the change in rate due to burnup over this period to be less than $1$\%.

The data presented in this paper is from three of our four cells. The excluded cell was found to have unacceptably large fluctuations in the energy calibration, of order $0.15$~MeV. These fluctuations are not fully understood, but are thought to be due to a failure of one or more of the gas-filled internal reflectors.

\subsection{Detector Stability}
\label{sec:stability}

Since we seek to observe a change in the antineutrino detection rate relative to that at the beginning of the reactor fuel cycle, the stability of our detector response is of paramount importance. We rely on the automated energy calibration to account for any time dependent variations in the gain of the scintillator/PMT/DAQ system, since an error in energy scale will result in a change in the threshold applied during event selection.

\begin{figure}[tb]
\centering
\includegraphics*[width=3in]{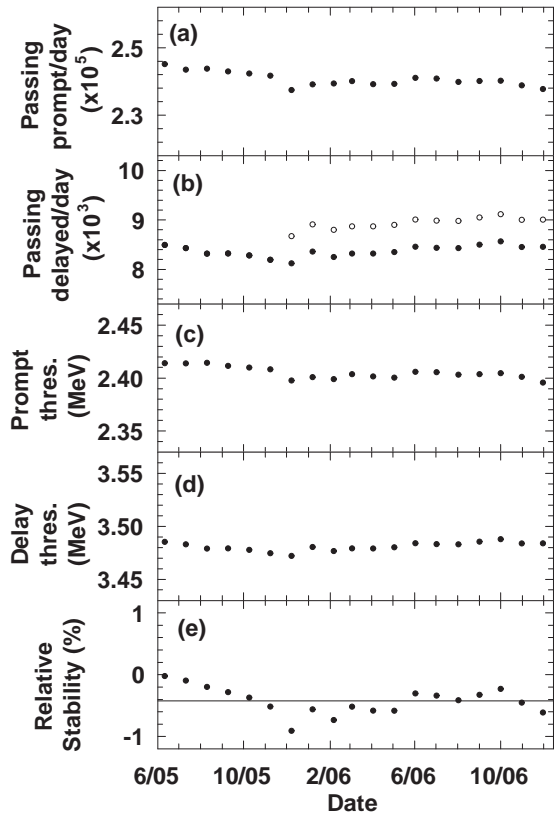}
\caption{The number of events that pass the prompt (a) and delayed (b) energy cuts during the data taking period. In (b) hollow circles after Nov. 9, 2005 denote the measured delayed rate, while solid circles denote the rate corrected for the movement of an AmBe source (the effect on the much larger prompt rate is negligible). The inferred prompt (c) and delayed (d) energy thresholds are determined from these single event rates. These thresholds allow us to estimate the relative stability of the antineutrino detection efficiency (e).}
\label{fig:stability}
\end{figure}

Examination of the variation of the number events passing the prompt and delayed energy cuts (Fig.~\ref{fig:stability}(a)\&(b)) allows us to estimate the effectiveness of this approach. Given that the background radioactivity that produces the vast majority of events in the detector is essentially constant, we would expect this number to remain constant if our single point calibration of the background gamma line were completely effective. Instead we observe small variations, implying that there is indeed some uncorrected drift in the energy calibration that we apply to the raw data.

To quantify the magnitude of drifts in the energy scale we measure the number of background events that pass the prompt and delayed energy cuts as a function of the energy threshold for those cuts. These event populations are rapidly decreasing exponential functions of the energy threshold value, so that small changes in threshold or errors in energy scale result in fairly large changes in the number of background events passing the cuts. By calculating the energy threshold that would have kept the number of passing events constant (Fig.~\ref{fig:stability}(c)\&(d)), we estimate the error in energy scale.

As can be seen in Fig.~\ref{fig:stability}, this error is less than $30~$keV over our $18$~month data taking period. Using the measured prompt and delayed energy distributions for antineutrino interactions~\cite{secondpaper}, we can estimate the effect of the remaining uncorrected gain shift on the measured antineutrino rate (Fig.~\ref{fig:stability}(e)). The effect is less than $1$\% over the data taking period, considerably smaller than the burnup effect we are attempting to observe, which is of order $10$\%.

There is one important exception to the assumption of a constant background radiation field used above, as noted in Fig.~\ref{fig:stability}(b). On November~9 of 2005, SONGS staff moved a $100~$mCi Americium-Beryllium (AmBe) neutron source ($\approx 300,000$ neutrons/s) from a position $25$~m around the circumference of the tendon gallery from the detector to one $20$~m away. At no time was there a clear line of sight between this source and the detector; however a small fraction of neutrons scattering from the annular gallery walls reach the detector. This resulted in an increase in the number of singles events passing the delayed (neutron capture) energy cut. Examination of the delayed cut singles rate with a 24~hour integration time verifies that there was a step increase on that date, which we therefore attribute to the movement of the AmBe source. The increase in the delayed singles rate on that date was $550$ per day ($\approx 3$ parts in $10^8$ of the neutrons emitted by the source during one day). In assessing the detector stability, we therefore subtract this step from dates following this event.

We are confident that the movement of the AmBe source to this new position did not cause a significant increase in correlated (antineutrino mimicking) background. An AmBe source could produce correlated background in two ways. First, a large fraction of neutrons emitted by an AmBe are accompanied by a $\approx 4$~MeV gamma ray. Detection of that gamma followed by capture of the correlated neutron would produce a correlated background event. The very small solid angle for this coincidence ($\approx (1/20^2)^2)$, combined with the requirement that both the neutron and gamma scatter off the gallery walls and traverse both the source and detector shielding unattenuated makes this possibility vanishingly small. Secondly, if an AmBe neutron reached the detector with sufficient energy to produce proton recoils above the prompt energy threshold, a correlated background event would result. This is again unlikely since the neutron would lose a substantial fraction of its energy while scattering along the gallery walls to reach the detector, as well as traversing at least $1$~m of hydrogenous material (source plus detector shielding). Furthermore, we observe no increase in the correlated event rate on the date that the AmBe source was moved. The reactor refueling outage that we observe began two months later, on January 3, 2006.

We note that this external source movement event is an example of actions that might occur during typical operations at power and, especially, research reactors. Careful examination of the singles and correlated events rates allows these changing background radiation fields to be accounted for. This also allows a means to detect spoofing attempts using neutron sources like AmBe or $^{252}$Cf that can produce correlated events. The addition of small fast and thermal neutron detectors could further strengthen the robustness of the system against such background variations or spoofing attempts.

We can also use correlated backgrounds as another estimator of detector stability. This background is primarily due to spallation neutrons that can  mimic the antineutrino double coincidence signal by depositing kinetic energy in the scintillator before thermalizing and being captured on Gd. In our typical analysis, we remove many of these events by vetoing the detector for a period of $100~\mu$s after the passage of a muon. Instead, measuring the rate of muon correlated events provides a measure of the stability of the detector response for antineutrino like events. Events within $20~\mu$s of a muon are vetoed in hardware, so we cannot access them. We can, however, apply this test to events in the range $20$~to~$100~\mu$s.

\begin{figure}[tb]
\centering
\includegraphics*[width=3in]{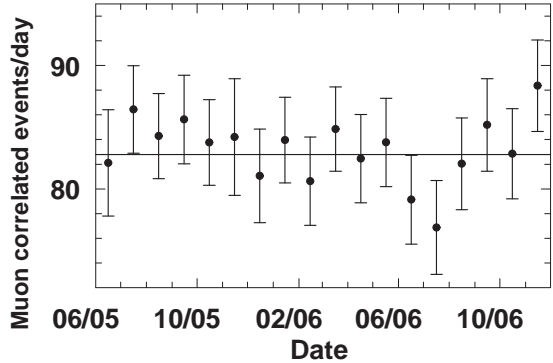}
\caption{Muon correlated event rate measured in the time period $20$~to~$100~\mu$s after the passage of a muon through the veto shield. A fit with slope constrained to zero is in good agreement with the data.}
\label{fig:stability2}
\end{figure}

Using the same statistical method as that employed to extract the antineutrino signal, we examine the distribution of double coincidence events that pass our analysis cuts, as a function of time after last muon or acquisition trigger. This distribution has two components, an exponential decay with a time constant equal to the inverse of the sum of muon and acquisition trigger rates (due to random coincidences between the event pair that passes the selection cuts and the preceding muon or trigger), and a component at small times correlated with the passage of the last muon (Fig.~$10$ of \cite{secondpaper}). We then subtract an exponential fit to the random coincidence background to extract an estimate of the number of muon correlated events that pass our cuts. In Fig.~\ref{fig:stability2} we have plotted these events in monthly bins. The distribution is well fit by a linear function with a slope constrained to zero ($\chi^{2}~=~13.5$, $16~d.o.f$). While this analysis is independent of the antineutrino flux, it is vulnerable to long term variations in the flux of cosmic ray muons, although this is expected to be less than $2\%$ \cite{muon_stab}. The analysis shows that the variability in the detector response is small for this class of antineutrino-like events, in comparison to the variations observed in the actual antineutrino detection rate (Sec.~\ref{sec:burnup}).

\section{Observation of Fuel Evolution}
\label{sec:burnup}

\begin{figure}[tb]
\centering
\includegraphics*[width=3in]{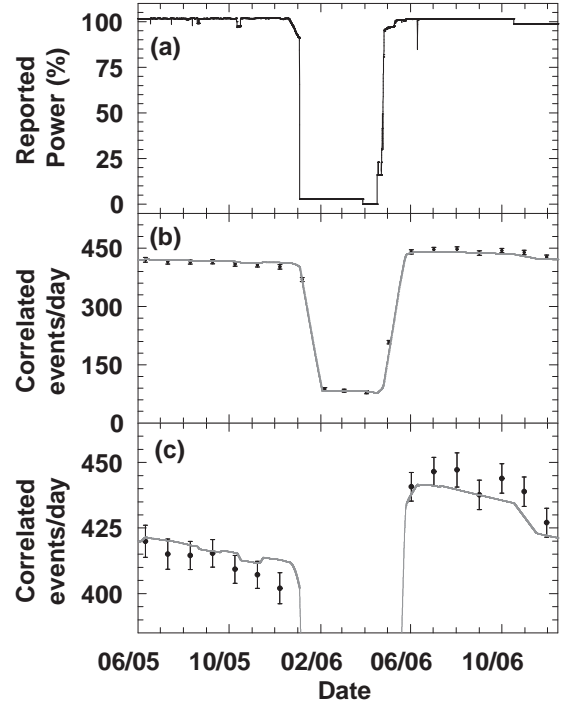}
\caption{(a) The reactor power during this same time period as reported by the operator. (b) Observed correlated event rate from the three detector cells used in this analysis (points) and predicted rate based upon a rolling average of the burnup model (solid line). (c) As in (b) but with expanded $y$-axis to better demonstrate the change in the measured rate after the refueling outage.} \label{fig:dataset}
\end{figure}

The power history of the SONGS Unit~2 reactor during the data taking period is shown in Fig.~\ref{fig:dataset}a. This plot includes the small contribution to the antineutrino rate coming from the SONGS Unit~3 reactor. Both power levels are scaled with their distance from the detector to the two active reactor cores. It can be seen that SONGS Unit~2 operated at full power for the last half of 2005 before shutting down for a refueling outage in January of 2006. The outage lasted for almost $4$~months, after which Unit~2 returned to only $99$\% of full rated power -- several tubes in the reactor's steam generators had to be plugged due to aging, reducing the generating capacity.

It is also interesting to note that there was a two week period in April 2006 during which both Unit~2 and Unit~3 were shutdown. During this time the correlated event rate is due only to background -- the ability to measure the background rate only may prove to be a valuable feature for planned experiments to measure the third neutrino mixing angle (see, for example,~\cite{DCLOI}). Experiments of this sort carried out at a plant with no more than two reactors have a reasonable chance of being able to make such a pure background measurement, as demonstrated here.

The correlated event rate, averaged for $30$~days, measured by the $3$~fully operational detector cells is plotted in Fig.~\ref{fig:dataset}b. The correlated event rate clearly tracks changes in reactor power that are long compared to the averaging time; for instance, the reactor outage is clearly visible. The reactor off period allows us to determine the correlated background rate, which in this instance includes the antineutrino contribution from Unit~$3$. During the $85$~days when Unit~$2$ was off, but Unit~$3$ was on, a correlated background rate of $83.7\pm2.2$ events/day was measured.

Also shown in Fig.~\ref{fig:dataset}b is our prediction of the correlated event rate based upon a $30$~day rolling average of Eq.~\ref{eq:nu_d_rate2}, using $k(t)$ determined by our burnup model (Sec.~\ref{sec:reactorsim}). Additionally, two parameters are determined from the data to produce the prediction; the correlated background rate, and a factor that accounts for the detector efficiency whose determination is described below.

In Fig.~\ref{fig:dataset}c, the $y$-axis is expanded to better display the correlated event rate when the reactor is operated at full power. We observe a clear decrease in rate, consistent with that expected . Compared with the rate just before the reactor shutdown in December 2005, a rate increase is clearly visible following reactor turn-on in June 2006. This results from the large exchange of fissile material that occurred during the reactor refueling. At the end of cycle (December~2005) the reactor core contains, and is burning, a large amount of $^{239}$Pu, and thus produces fewer antineutrinos. During refueling, approximately one third of the reactor fuel, containing $\approx 250$~kg of $^{239}$Pu, was removed and replaced with fresh fuel containing $\approx 1.5$~tons of $^{235}$U.

\begin{figure}[tb]
\centering
\includegraphics*[width=3in]{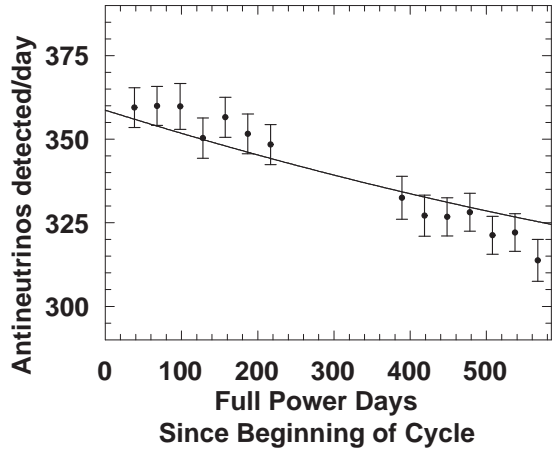}
\caption{Observed antineutrino rate plotted as function of the number of days the reactor has operated at full power ($\propto$ fuel burnup).
The solid curve is a one parameter fit of the rate predicted by our reactor
simulation to the data ($\chi^{2}~=~13.3$,
$13~d.o.f$). } \label{fig:burnup}
\end{figure}

Since SONGS Unit~2 has been operational for many fuel cycles, its core evolution will have reached equilibrium, i.e., the core evolution will be similar from cycle to cycle. Thus we can take the data that we have collected for the beginning and end of different cycles and plot them as though they belonged to the same cycle (Fig.~\ref{fig:burnup}). Here we have subtracted the correlated background rate, and have scaled the rates to account for small variations in reactor power ($<2\%$).

There is a clear decrease in the detected antineutrino rate as the cumulative power produced increases. The observed reduction is in good agreement with the prediction based upon our reactor simulations -- the solid curve in Fig.~\ref{fig:burnup} is a one parameter fit of the burnup model described in Sec.~\ref{sec:reactorsim} to the data ($\chi^{2}~=~13.3$, $13~d.o.f$). Furthermore, the observed reduction in detected rate is not due to any instability in our detector -- the observed reduction is of order $10$\%, while the detector efficiency and  correlated background rate are stable at the level of $\approx 1$\%~(Sec.~\ref{sec:stability}).

The SONGS1 detector is therefore sensitive to both the slow change in antineutrino detection rate as the reactor fuel evolves, and the step change in rate that occurs if fuel is replaced.

Assigning a sensitivity in terms of fissile mass is difficult as these changes in detection rate do not directly relate to a particular change in fuel isotopics. To provide an indication of our sensitivity to sudden changes in fuel isotopics, e.g. the undeclared replacement of spent fuel, we have performed a Monte Carlo study to determine the change in the antineutrino detection rate that could be reliably observed by the SONGS1 detector.

\begin{table}
\caption{The change in daily antineutrino ($\bar{\nu}$) detection rate that could be reliably observed by SONGS1 for a range of observational confidence parameters. The relationship between $\bar{\nu}$ detection rate and fissile material quantity is scenario dependent; in the fuel exchange (refueling) observed in this work, replacement of $\approx 250$~kg of $^{239}$Pu with $\approx 1.5$~tons of $^{235}$U resulted in a detection rate change of $\approx 35$~$\bar{\nu}$ per day. The False Positive Rate is how often statistical fluctuations produce an indication of a step change in the $\bar{\nu}$ detection rate of the magnitude given in the table, in the absence of any actual change in detection rate. The False Negative Rate (FNR) is how often a true step change in the $\bar{\nu}$ detection rate of the magnitude given in the table is masked by statistical fluctuations.} \label{tab:sen_mc}
\begin{tabular}{c c c} \hline
 False Positive Rate~~~~&\multicolumn{2}{c}{Change in $\bar{\nu}$ detection rate}\\
 &FNR = 1 in 10 &FNR = 5 in 10 \\ \hline
1 in 10$^3$ & ~~11.6&  9.1 \\
1 in 10$^4$ & ~~12.9&  10.2\\
1 in 10$^5$ & ~~14.2&  11.5\\
\hline
\end{tabular}
\end{table}

In this study, the SONGS1 antineutrino detection rate was simulated throughout a 600 day fuel cycle. At day~300 in the cycle, the antineutrino detection rate is increased, as it would if Pu laden fuel were exchanged for fresh fuel. For each value of the step change in rate $100,000$~simulated experiments were conducted, sampling from gaussian distributions to represent the systematic uncertainty in our knowledge of thermal power ($0.5\%$) and the magnitude of the burnup effect ($3\%$), as well as the statistical fluctuation of the antineutrino detection rate, and a $\chi^{2}$ distribution generated. Comparison of these distributions to that generated with zero step allows us to estimate the change in antineutrino detection rate required to observe such a change (Table~\ref{tab:sen_mc}). For comparison, the step change that occurs at refuelling, when $\approx 250$~kg of $^{239}$Pu is replaced with $\approx 1.5$~tons of $^{235}$U, is $\approx 35$~counts per day.

\section{Conclusion}
\label{sec:conclusion}

The SONGS1 data demonstrate a clear correlation between the changes in the reactor antineutrino emission rate and the evolution of the reactor power and fissile inventory. While the sensitivity differs according to the specific fuel loading and operating history, our data and simulations show that changes in antineutrino rate throughout a fuel cycle, including both rate ``slope'' and ``step'' information, can be used in conjunction with power declarations to constrain the amount of $^{239}$Pu being produced or removed at the level of about $100$~kg.

High-statistics antineutrino rate measurements of the kind presented here can be used to quantitatively verify operator inventory and operating history declarations, complementing the containment and surveillance based approach that prevails in the current IAEA reactor safeguards regime. Comparison of the antineutrino rate changes from fuel cycle to fuel cycle also provides independent verification that the reactor is being run in a consistent fashion across cycles. Since in this case only self-consistency of the antineutrino measurements across cycles is demanded, such an approach reduces dependence on operator declarations.

The SONGS1 deployment has also demonstrated that an antineutrino detector can be operated stably and without interruption or maintenance for many months to a year or longer, using background gamma rays as the sole means of calibration, and with remote real-time retrieval of data and detector state-of health information. With appropriate protection against tampering, this makes remote quantitative verification of reactor operations possible, and may thereby allow a reduction in the frequency of site visits by IAEA or other inspectors.

Further work must be done to study specific scenarios, and to consider safeguards tasks of particular interest to the IAEA, such as the study of different reactor types and power ratings. In terms of detector development, further improvements in precision and ease of operation are being pursued through iterations on the current detector design. Ultimately, complete independence from declarations may be made possible through measurements of the antineutrino energy spectrum, using a somewhat more sophisticated detector.

\begin{acknowledgments}
We are deeply indebted to the management and staff of the San Onofre Nuclear Generating Station for site access and the excellent support they have provided us. N.~Madden, D.~Carr, J.~Gollnick, J.~S.~Brennan, and A.~Salmi made essential contributions to the electronic and mechanical design and construction of the detector. We are particularly grateful to F.~Boehm for providing us with the liquid scintillator used in this detector. We also acknowledge important early input to the analysis by Mark Cunningham.

LLNL-JRNL-401959.

This work was performed under the auspices of the U.S. Department of Energy by Lawrence Livermore National Laboratory in part under Contract W-7405-Eng-48 and in part under Contract DE-AC52-07NA27344.

SAND Number: 2008-1587J.

Sandia National Laboratories is a multiprogram laboratory operated by Sandia Corporation, a Lockheed Martin Company, for the United States Department of Energy under contract DE-AC04-94AL85000.
\end{acknowledgments}


\end{document}